\documentclass[sigconf]{acmart}

\usepackage{bytefield}
\usepackage{rotating}
\usepackage{balance}
\usepackage{pifont}
\newcommand{\cmark}{\ding{51}}
\newcommand{\xmark}{\ding{55}}
\usepackage{subfigure}
\usepackage{alltt}
\usepackage{hyperref}

\begin{document}
\settopmatter{printacmref=true}
\fancyhead{}

\title{TOUCAN}
\subtitle{A proTocol tO secUre Controller Area Network}

\author{Giampaolo Bella, Pietro Biondi}

\affiliation{%
  \institution{Dipartimento di Matematica e Informatica\\ Universit\`a di Catania}
}
\email{giamp@dmi.unict.it, pietro.biondi94@gmail.com}

\author{Gianpiero Costantino, Ilaria Matteucci}
\affiliation{%
  \institution{Istituto di Informatica e Telematica \\ Consiglio Nazionale delle Ricerche}
  }
\email{gianpiero.costantino@iit.cnr.it, ilaria.matteucci@iit.cnr.it}

\begin{abstract}

Modern cars are no longer purely mechanical devices but shelter so much digital technology that they resemble a network of computers. Electronic Control Units (ECUs) need to exchange a large amount of data for the various functions of the car to work, and such data must be made secure if we want those functions to work as intended despite malicious activity by attackers.
TOUCAN is a new security protocol designed to be secure and at the same time both CAN and AUTOSAR compliant. It achieves security in terms of authenticity, integrity and confidentiality, yet without the need to upgrade (the hardware of) existing ECUs or enrich the network with novel components. The overhead is tiny, namely a reduction of the size of the Data field of a frame. A prototype implementation exhibits promising performance on a STM32F407Discovery board.
\end{abstract}


\maketitle

%------------------------------------------------------------------------------
\section{Introduction}
\label{sect:introduction}
Modern vehicles abound with Electronic Control Units (ECUs) that need to speak with each other. Components such as airbags, power doors, electric mirrors need to interconnect and communicate to ensure the smooth and synergistic functioning of all. They adopt a binary language and form an in-vehicle network that must be precisely regulated. This was the aim for the inception of ``Controller Area Network'' protocol, also known as \textit{CAN bus}, which dates back to 1983 with Bosch \cite{summacan} and is widespread today. It is standardised in ISO 11898-1:2015~\cite{isocan} as a simple protocol based on two bus lines. The layman might conjecture the existence of the CAN when his car is diagnosed in a lapse at the mechanic's through a laptop connected to the 16-pin OBDII/EOBD port of the car. The CAN protocol runs over two of the pins of such port.

The CAN bus is not meant to be secure. It signifies that it was designed in the assumption that it would execute in a friendly environment, with participants sharing the common goal of the smoothest possible functioning of the host car. Cybersecurity researchers are well aware that overly optimistic assumptions are deemed to be broken eventually. The current automotive landscape makes no exception and clearly breaks the assumption of a friendly execution environment for several reasons.
One is that cars' functioning is closely related to passenger safety, hence potential threats may be ill-driven towards harming people. Another one is that cars are becoming more and more tightly interconnected to the external world, not only by GPS, but also by 4G and by dedicated connections such as for \emph{e-call} boxes \cite{ecall}. They hold a variety of driver's (and, progressively, passengers') data, such as driving style or significant episodes, environmental data acquired through various sensors and cameras, as well as data received from the passengers' smartphones. Where biometric techniques are used to authenticate the driver, cars even treat sensitive data. All such data is obviously appealing for profiling and marketing reasons, e.g., for insurance companies.

Therefore, awareness is growing on the need to secure in-vehicle communication, but this goal is daunting: whatever novel technology should be tested on the large scale, should not be prohibitively expensive and, above all, should not clash with efficiency of the communication, because this is tightly related to the latency of the various devices of a car, hence again to passenger safety. 

This paper presents TOUCAN, a protocol to secure the CAN bus against an active eavesdropper. The protocol enjoys backward compatibility with existing standards and requires no hardware upgrade but solely a firmware update to implement TOUCAN.
TOUCAN uses a fast hashing algorithm, Chaskey~\cite{chaskey14}, to provide authenticity and integrity of the payload of a frame, and AES-128 encryption for confidentiality. The TOUCAN frame complies with the CAN standard hence its Data field is of 64 bits, although it carries an actual payload of 40 bits because its hash is carried by the remaining 24 bits of the field.

The prototype implementation of both hashing and encryption algorithms used in TOUCAN has been tested on inexpensive hardware.
In particular, our test-bed is composed by a STM32F407 Discovery board in which we deploy programs computing Chaskey hash values as well as AES-128 cyphertexts, as we shall see. The runtimes we measured were promising, a result that also makes TOUCAN a potentially good candidate for adoption as an AUTOSAR in-vehicle 
security protocol. 

This paper continues with the details of the TOUCAN protocol (\S\ref{sec:algorithm}), its implementation and performances (\S\ref{sec:impl}). Finally, it draws the relevant conclusions (\S\ref{sec:concl}). A primer on the CAN bus is available in Appendix (\S\ref{sec:can}).
\section{The TOUCAN protocol}\label{sec:algorithm}
The mentioned lack of fully-fledged security measures at the level of CAN bus communication motivates the need for a security protocol. TOUCAN aims at securing communication among Electronic Control Units (ECUs) interconnected via CAN. It is not the first security protocol in this area with this aim but, as we shall see below (\S\ref{sec:sota}), it conjugates a number of useful features.

TOUCAN assumes a realistic threat model. An active attacker has exploited some connections to the vehicle and broken any boundary protection that may be in place. Therefore, the attacker can eavesdrop all CAN bus traffic, namely intercept, record and modify frames to maliciously reuse them, e.g., breaking the vehicle, steering the wheel and so on. In practice, this may take place remotely, through a 4G connection, possibly by exploiting some vulnerability of (any of the levels of) the software executed in the car. Another attack vector may derive from physical access to the OBDII port while the car is unattended~\cite{ODBII1,ODBII2}.

The current level of development of TOUCAN assumes that the relevant cryptographic material is available to all ECUs. Therefore, control units share secure keys that can be used to hash or encrypt their communication. There exists literature in this area~\cite{PnS2015}, which we are working to customise and integrate with our protocol.

The design of TOUCAN is loosely inspired to a recent protocol by Dariz et al.~\cite{Dariz2018,darizmts}. That protocol prescribes the use of Message Authentication Code (MAC) and encryption on the payload of a CAN frame to ensure security in terms of authenticity, integrity, and confidentiality. The protocol is not implemented, as the main goal is to mathematically assess how the overhead of the security measures may impact the safety of a vehicle. 

 An essential by design requirement of TOUCAN is to comply with existing standards in this area. First of all, it is CAN 2.0 compliant, which means that it can be executed on existing ECUs and infrastructures running CAN. It is also compliant with the AUTOSAR standard on ``On-board Secure Communication''~\cite{autosar422}, as it is designed to satisfy one of the specific \emph{profiles} defined by the standard~\cite{autosar431}. A profile is an assignment of values to four configuration parameters: algorithm for encryption, length of freshness value (parameter \texttt{SecOCFreshnessValueLength} holding the length in bits of the freshness value), length of truncated freshness value (parameter \texttt{SecOCFreshnessValueTXLength} denoting the length in bits of the freshness value to be included in the payload of secured messages)  and length of truncated MAC. The profile that TOUCAN in its current version adopts is reported in Table~\ref{tab:profile2}, but also the other profiles are worthy of future experimentation.
\begin{table}[ht]
\centering
\scalebox{0.8}{
\begin{tabular}{l c}
\hline\hline
Parameter & Configuration Value \\ [0.5ex] % inserts table %heading
\hline
Algorithm&CMAC/AES-128 \\
Length of Freshness Value\\ (parameter \texttt{SecOCFreshnessValueLength})& 0 bit\\
Length of truncated Freshness Value \\(parameter \texttt{SecOCFreshnessValueTXLength})& 0 bit\\
Length of truncated MAC \\(parameter \texttt{SecOCAuthInfoTXLength})& 24 bits\\ [1ex]
\hline
\end{tabular}}
\caption{The AUTOSAR profile that TOUCAN conforms to}
\label{tab:profile2}
\end{table}
TOUCAN leverages the Chaskey MAC, which is  a very efficient permutation-based MAC algorithm based on Addition-Rotation-XOR (ARX) with some useful features. Among the main ones is its robustness under tag truncation. In terms of design, Chaskey does not require a nonce and is well suited for 32-bit micro-controllers.

According to the requirement of the mentioned AUTOSAR profile, we are allowed a MAC field of 24 bits, which implies that we can rely on a payload of 40 bits in order to fit both of them in the Data field of a frame. By using a 64-bit Data field, TOUCAN is CAN compliant. Our protocol prescribes the use of Chaskey to compute the MAC, precisely the application of Chaskey to the payload of each frame, then the truncation of the outcome to 24 bits to fit in the rest of the Data field. The truncation resistance of Chaskey is a clear advantage here. TOUCAN conveys payload authenticity and integrity because each Chaskey application requires a 128-bit key as additional parameter, and such a key is assumed to be correctly managed.
\begin{figure*}[t]
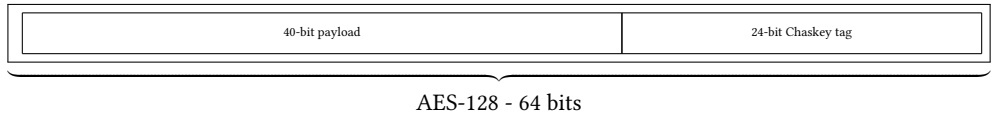

	\begin{bytefield}[endianness=little,bitwidth=.63em]{64}
		%\bitheader{0-63}\\
		$
		\underbrace{
			\framebox{
				\bitbox{40}{\tiny 40-bit payload}
				\bitbox{24}{\tiny 24-bit Chaskey tag}}
		}_
		{\text{\normalsize AES-128 - 64 bits}}
		$
	\end{bytefield}
	\caption{The TOUCAN Data field}\label{fig:securecanframe}
\end{figure*} 
Once TOUCAN forms the hashed message, it encrypts the entire Data field by using the Advanced Encryption Standard with a 128-bit key (AES-128), according to AUTOSAR profile. AES is based on a design principle: a substitution-permutation network, hence is efficient for both sw/hw execution. Encryption preserves size, hence the TOUCAN Data field is as depicted in Figure \ref{fig:securecanframe}.

\subsection{Design evaluation}
TOUCAN does not require additional hardware components and is backward compatible, but this does not come without limitations.

As it is well known, the CAN bus has limited bandwidth, short data frames and a publisher-subscriber broadcast architecture where new messages overwrite the older ones in the receiver's buffer.
Therefore, there are two possible approaches to secure CAN communications. One consists in increasing the number of CAN frames circulating on the network, specifically by sending more than one frame for each original CAN frame. This is typically due to the fact that the secured payload of the original frame must be fragmented over more frames to also carry a MAC. However, this solution has been found to potentially impact safety because it may increase the latency of the response time~\cite{darizmts}. The other approach is to reduce the payload carried per frame, as TOUCAN does. This decreases the number of messages that the car manufacturer can leverage to implement modern services based on communication among ECUs. Although we argue that a message space of $2^{40}$ is sufficient, this will have to be validated over time as more and more developed applications appear.
A MAC is considered robust when its size exceeds 64 bits, otherwise we must carry out a probability analysis of guessing a tag (finding counter-images) and of collisions (birthday attacks)~\cite{Dworkin05,chaskey14}.
\\
\textbf{Risk of guessing the tag.} According to~\cite{chaskey14}, the probability of constructing a forgery by guessing the tag is $2^{-tag\_len}$. Hence, in TOUCAN, being $tag\_len$ equal to 24, the probability of guessing a tag is $0,6*10^{-7}$.  \\
\textbf{Probability of tag collisions.} Referring to the definition in~\cite{birthday}, the collision probability depends on both the MAC length and the number of times the MAC is calculated. So, the boundary limit before collision is $2^{\frac{tag\_len}{2}}=2^{\frac{24}{2}}=2^{12}=4096$. 
\section{A prototype implementation}\label{sec:impl}
We developed a prototype implementation of the main components of TOUCAN, namely hashing and encryption, with very promising performance results. Our prototyping environment can be summarised as follows.

\begin{itemize}
\item The programming effort takes place on the ``Keil $\mu$Vision Version 5'' IDE, developed by ARM, which comes with a powerful debug environment, and supports multiple project management.
\item 
The testbed is an ``STM32F407 Discovery'' board, controlled by an ARM Cortex M4 processor and equipped with LED and push-buttons, and connected via USB to the PC.
\item
The code is deployed to the board by means of ST's proprietary software ``STM32CubeMX'' \cite{stcube}, which comes with a useful graphical interface. 
\item
The performance evaluation phase is managed on a VirtualBox virtual machine running on a Windows 10 host PC with 4GB of RAM and 50GB of hard disc. 
\end{itemize}

\subsection{Hashing}
A prototype implementation of Chaskey is already available, hence we take advantages of the source code that can be found on GitHub \cite{chaskeylib}. However, it cannot be used as is, and additional steps are needed to execute it on our STM board.

\begin{figure*}[ht]
  \centering
  \includegraphics[scale=0.5]{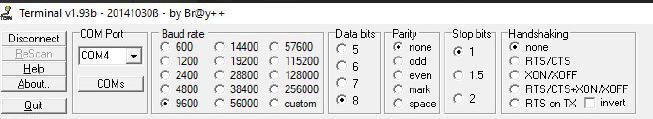}
  \caption{Configuration of USART-to-USB interface}\label{fig:settings_terminal_usart}
\end{figure*}

\paragraph{Implementation of the USART communication.} 
A \emph{Universal Synchronous Asynchronous Receiver/Transmitter} (USART) can be used to setup a serial communication between the PC and the board. First, the developer has to set the pin of the board on which it transmits and, in terms of software, some libraries and data structures to establish the communication. The STM32CubeMX offers good support to generate and initialise the relevant C code to flash the board. The PB7 pin on the board must then be initialised with value USART1 TX to transmit and the PB6 pin with value USART1 RX to receive. Another parameter to set is the baud rate, which can be up to 9600. All these parameters can be set through the interface as shown in Figure \ref{fig:settings_terminal_usart}. 

Once all parameters are configured, the C code is generated and flashed to the board via the USART-to-USB interface, which is activated in listening mode to receive messages.
A buffer for transmission and one for reception are initialised:
\small{\begin{alltt}
char* bufftr;
uint8\_t buffrec[10];
\end{alltt}
}
\normalsize
Notably, the USART-to-USB interface is established by executing the HAL (\emph{Hardware Abstraction Layer}) API in a never-ending while cicle in such a way that the communication is maintained constantly:
\small{\begin{alltt}
HAL\_UART\_Transmit\_IT(\&huart1, (uint8\_t *)bufftr, 8 );
HAL\_UART\_Receive\_IT(\&huart1, buffrec, 10);
\end{alltt}}
\normalsize
\noindent where the function \texttt{HAL\_UART\_Transmit\_IT} regulates the transmission of messages. It takes as input the USART address \texttt{huart1}, the buffer to be transmitted and its size. The function \texttt{HAL\_UART\_Receive \_IT} is similar to the previous one with the difference that the received message is inserted in the buffer of received messages.

\paragraph{Execution.}
Once the USART interface is configured and activated in listening mode as explained above, we built a project using the Chaskey API \texttt{chas\_mac}. This is compiled and flashed to the board. At execution time, the message \texttt{OK} is received if the execution ends correctly, or \texttt{FAIL} otherwise. Runtimes are given below.
\subsection{Encryption}
To execute encryption with AES-128 on our STM32F407 Discovery board, we leverage the ST library named ``MDK-ARM'', which can be found as CryptoSTM32F4/Projects/STM32F401
RE-Nucleo/AES/AES-128\_CTR/MDK-ARM~\cite{STMlib}. Because the library is implemented for the STM32F401RE-Nucleo board, we had to make a number of modifications, omitted here, to a number of initialisation functions for the library to also run on our board.
\begin{figure*}[ht]
    \centering
    \begin{subfigure}
        \centering
        \includegraphics[height=1.5in]{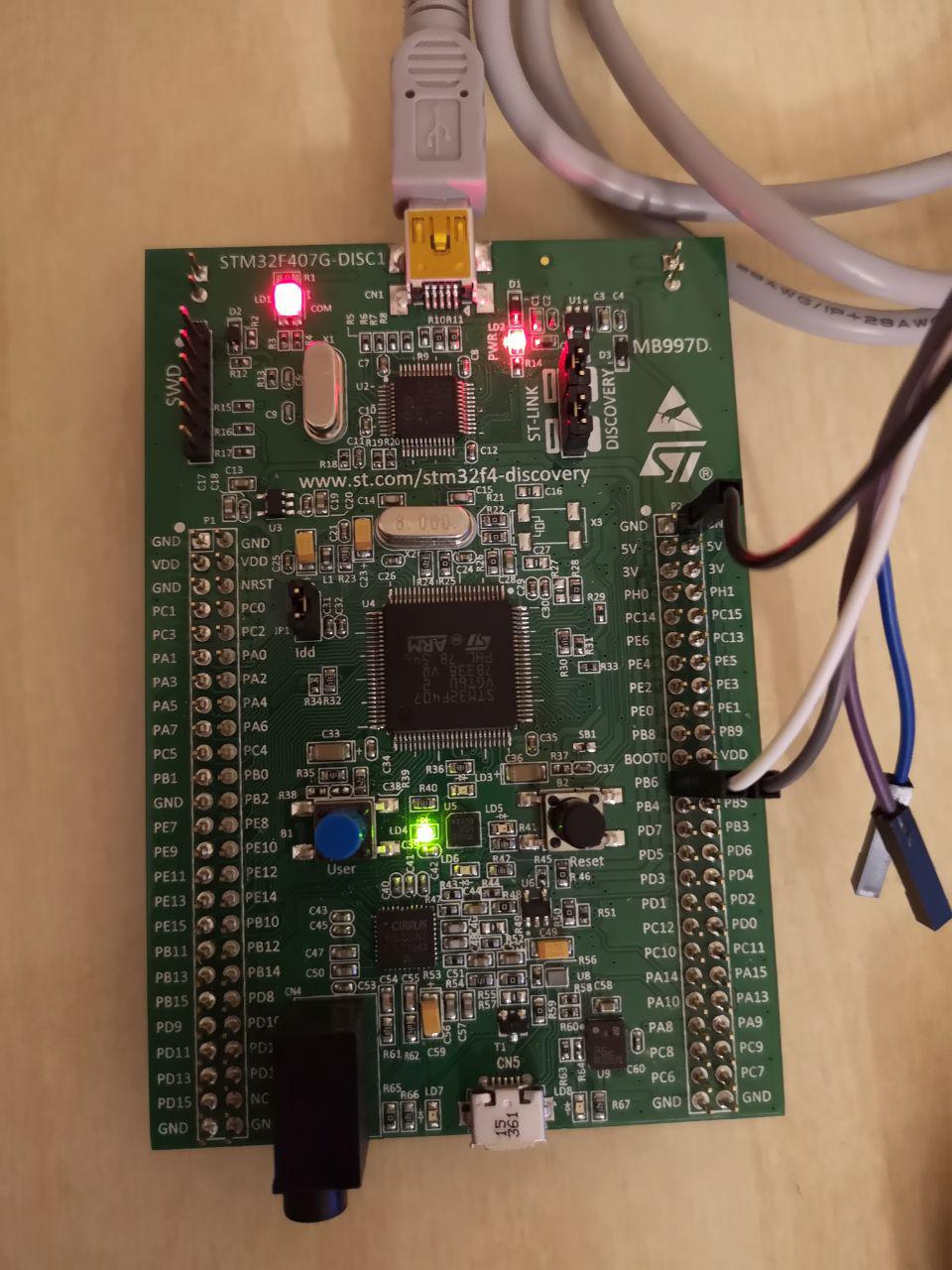}
        %\caption{The execution ends correctly}
    \end{subfigure}%
\qquad
    \begin{subfigure}
        \centering
        \includegraphics[height=1.5in]{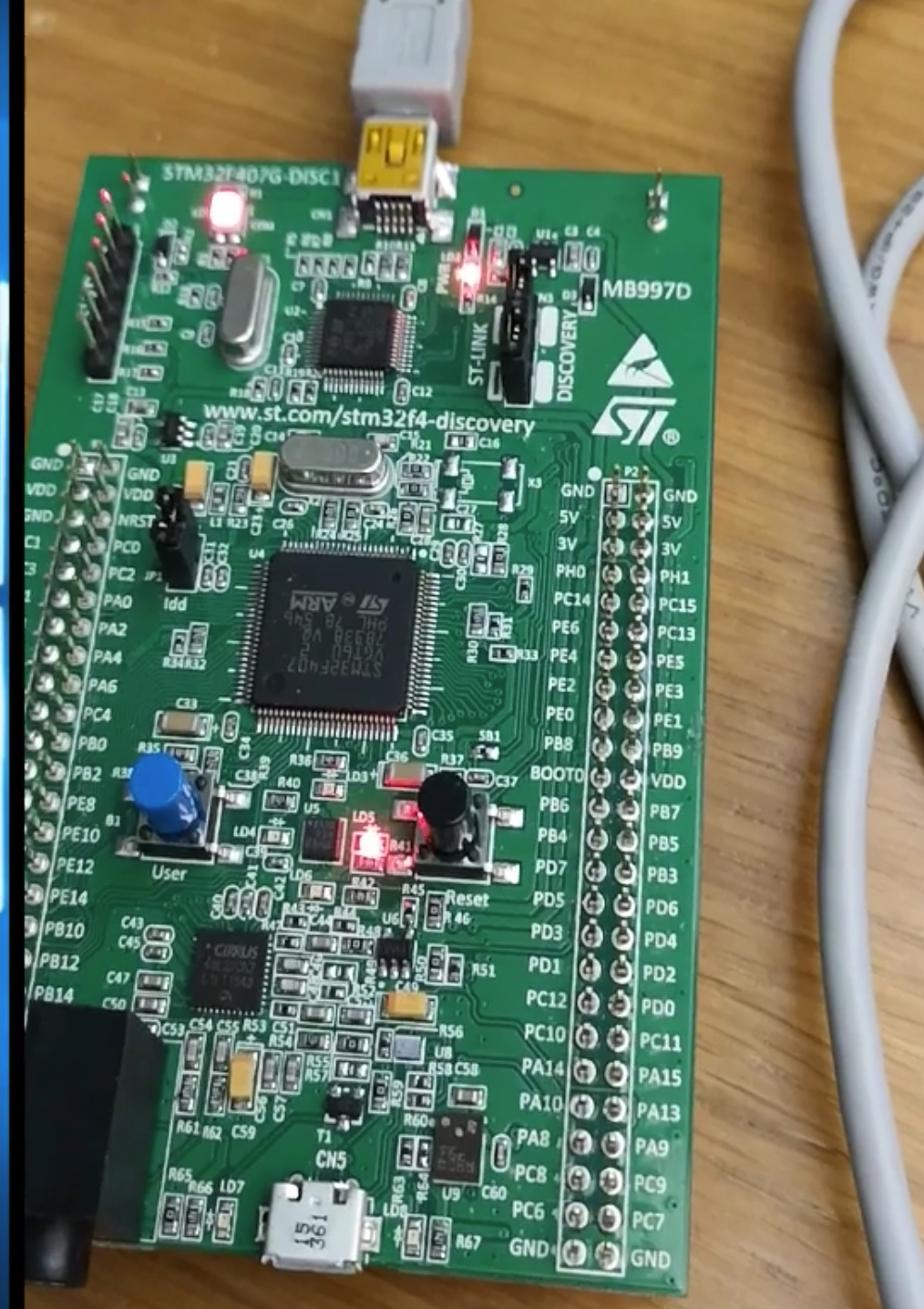}
          %\caption{The execution ends with an error}
    \end{subfigure}
     \caption{The outputs of our program running on a STM32F407 Discovery board}\label{fig:board}
\end{figure*}
We program the board so that it takes as parameters a 64-bit specimen cypher-text, a 64-bit clear-text and an AES-128 cryptographic key. The board encrypts the clear-text with the key and compares the outcome to the specimen cypher-text If they match, it turns the green led on (Figure~\ref{fig:board}, left), otherwise the red one (Figure~\ref{fig:board}, right).

We compute a cypher-text by encrypting a random payload, representing the Data field of a TOUCAN frame with a random AES-128 key, using an online tool. That cypher-text can be given as input to the board to verify whether also the board can compute it correctly. During our tests, we observed that everytime the board executed on a specimen cypher-text and on the same plaintext-key pair using to compute the specimen, it would then switch on the green light, otherwise the red light. It then seems fair to conclude that our board can correctly execute AES-128 encryption.
\subsection{Performance evaluation}
Table~\ref{tab:performance} reports our experiments on executing Chaskey and AES-128 on a payload of 64 bits representing a TOUCAN frame. They are carried out on an STM32F407 Discovery board, with an ARM Cortex M4 processor with promising runtimes. % are promising.
\begin{table}[h] 
\centering
\scalebox{0.9}{
\begin{tabular}{|c|c|c|}
	\hline 
	\textbf{Algorithm} & \textbf{Board Speed} [MHz] & \textbf{Time} [$\mu$s]\\ 
	\hline
	\hline  
	AES 128 bit - len 8 (byte) & 84 & 11,65\\ 
	\hline 
	Chaskey & 84 & 11,90\\ 
	\hline 
\end{tabular} 
}
\caption{Runtimes for the main components of TOUCAN}\label{tab:performance}
\end{table}
%\vspace{-8mm}
\section{Comparison with the related work}\label{sec:sota}
This section reports the main entries in the state of the art on securing in-vehicle communication. 

One of the earlier protocols is named CANAuth~\cite{canauth11} and dates back to 2011. 
It is based on CAN+~\cite{canplus09}, which is an extension of the basic CAN protocol in which the data rate is extended in such a way that more bytes can be sent (up to 16 CAN+ bytes for each CAN byte) in the same frame. The drawback of this protocol is that it requires to change the transceivers, which must be more powerful to manage the CAN+ data rate. This implies that using CANAuth has an impact on hardware, which must be upgraded.

In 2012, LCAP~\cite{LCAP12} was advanced with the aim of guaranteeing message authentication, resistance to replay attacks, and backward compatibility at the same time. 
As CANAuth, it is based on some out-of-band protocol like CAN+. The main drawback is the use of broadcast-based authentication, which increases the traffic in a way directly proportional to the number of nodes in the network.

In the same year, Hartkopp \emph{et.al.} proposed MaCAN~\cite{MaCAN12}. It is a centralized authentication protocol based on broadcast-based authentication, so it requires CAN+ or CAN FD. However, the same protocol was found to be flawed \cite{10.1007/978-3-319-10181-1_15}.

Also of 2012 is Libra-CAN by Groza \emph{et al.}~\cite{libra12}, a protocol based on a MAC calculated using MD5. Its main drawbacks are hight bandwidth and the introduction of hardware capable of understanding and manage the new frame format: Libra-CAN protocol is based on CAN+ instead of on CAN.

The relevant contributions of 2012 also include a mechanism by Lin and Sangiovanni-Vincentelli to prevent masquerading and replay at the cost of reasonable communication overhead~\cite{lin2012cyber}.

In 2014, CaCAN~\cite{CaCAN14} was introduced to ensure authentication and integrity of CAN messages but not confidentiality. In fact, the protocol consists on a key distribution phase inherited by existing protocols, and on the authentication and integration phase in which a MAC of the payload is calculated. For each message, three messages are sent, one with the payload in the clear, one with both payload and MAC and one with MAC as a payload. The protocol needs a new component to be inserted in the vehicular network in order to act as a monitoring node. Messages are not sent in broadcast but on a peer-to-peer base.

\begin{table*}[th]
	\centering
	\scalebox{0.9}{
		\begin{tabular}{l | c |c | c | c | c| c| c |}
			
			&\rotatebox[origin=l]{90}{CANAuth~\cite{canauth11}}
			&\rotatebox[origin=l]{90}{MaCAN~\cite{MaCAN12}}
			&\rotatebox[origin=l]{90}{LCAP~\cite{LCAP12}}
			&\rotatebox[origin=l]{90}{Libra-CAN~\cite{libra12}}
			&\rotatebox[origin=l]{90}{CaCAN~\cite{CaCAN14}}
			&\rotatebox[origin=l]{90}{LeiA~\cite{RaduG16}}
			&\rotatebox[origin=l]{90}{TOUCAN}\\
			\hline
			
			F1. Standard CAN& \xmark & \xmark & \cmark  & \xmark & \cmark & \cmark & \cmark \\
			\hline
			F2. Frame rate equal to CAN's. & \xmark & \xmark  & \xmark  &\xmark   & \xmark & \xmark & \cmark\\
			\hline
			F3. Payload size not smaller than CAN's. & \xmark & \xmark  & \xmark  &\xmark  & \xmark  & \cmark & \xmark\\
			\hline
			F4. Standard AUTOSAR& \xmark & \xmark  & \xmark  &\xmark   & \xmark  & \cmark & \cmark\\
			\hline
			F5. No ECU hardware upgrade & \xmark & \xmark  & \cmark & \xmark  & \cmark & \cmark & \cmark \\
			\hline
			F6. No infrastructure upgrade & \cmark & \xmark  & \cmark & \cmark  &  \xmark & \cmark & \cmark \\
			\hline
			\hline
			& 1 & 0  & 3 & 1  &  2 & 5 & 5 \\
		\end{tabular}
	}
	\caption{Contrastive analysis of TOUCAN w.r.t. the related work}\label{table:contrastive}
\end{table*}

The most recent protocol is LeiA (2016)~\cite{RaduG16}. It reaches message authenticity by using a MAC. For each message, the protocol sends a message in plaintext and another one with the MAC of the message.
LeiA rests on a 29-bit message identifier, which is coherent with CAN 2.0B \cite{isocan}.

All these protocols present some pros and cons, hence recent work analyses the impact of introducing security on functional properties of vehicles. In this vein, Dariz \emph{et al.}~\cite{Dariz2018,darizmts} presented a trade-off analysis between security and safety when a security solution based on encryption is applied on CAN messages. The analysis is presented considering different attacker models, packet fragmentation issues and the residual probability of error of the combined scheme. Also Groza \emph{et al.}~\cite{Groza18} and Stabili \emph{et al.}~\cite{Stabili18} targeted the delicate relation between security and safety.

With this picture in mind, we propose TOUCAN a new protocol for securing in-vehicle communication. 
We introduce an evaluation criterion to address the state of the art in terms of compatibility with standards and upgrades required to the traditional in-vehicle infrastructure. Although it is not the only possible one, our criterion rests on six separate \emph{features} F1 \ldots F6 that are remarkable to the future uptake and industrial deployment of secure in-vehicle communication.

\begin{itemize}
\item[\textbf{F1.}] \textbf{Standard CAN.} This holds of a protocol when all fields of the frame, which the protocol defines, conform to size and contents as they are specified by the CAN standard \cite{isocan}.
\item[\textbf{F2.}] \textbf{Frame rate equal to CAN's.} This is true for a protocol that does not need to send more frames than CAN does. For example, LeiA doubles the rate, and clearly falsifies this feature as well as any other protocol that requires the ISO-TP~\cite{ISO2016} fragmentation.
\item[\textbf{F3.}] \textbf{Payload size not smaller than CAN's.} This holds of a protocol that preserves the standard CAN size of 64 bits for the payload size. For example, TOUCAN falsifies this feature because it only relies on 40 bits of actual payload.
\item[\textbf{F4.}] \textbf{Standard AUTOSAR.} This holds of a protocol that conforms to the prescription of the latest AUTOSAR standard \cite{autosar422,autosar431}. Note that, AUTOSAR profiles have been recently introduced in the standard (2014).
\item[\textbf{F5.}] \textbf{No ECU hardware upgrade.} This holds of a protocol when it requires no upgrade to the ECUs that can run the CAN protocol, hence no additional features or computational power are needed for the units.
\item[\textbf{F6.}] \textbf{No infrastructure upgrade.} This is similar to the previous feature but concerns the network and the overall infrastructure that supports the protocol. Therefore it is true for a protocol that executes on the same network that underlies the CAN, without additional, dedicated nodes.
\end{itemize}
Table \ref{table:contrastive} adopts this criterion to represent a contrastive analysis of the main entries in the related work with respect to all six features. Notably, no protocol ticks all features, but LeiA and TOUCAN are the only protocols that are both CAN and AUTOSAR compliant and, at the same time, require no upgrade to each ECU, or network augmentation with additional components. However, the two protocols have alternate features F2 and F3. While LeiA keeps the CAN payload size of 64 bits, it doubles each frame, a feature that may produce some safety concerns, as discussed elsewhere \cite{darizmts}. 

By contrast, TOUCAN resolves this issue by carving out bits from the payload size, which reduces to 40 bits.
We argue that it will be possible to choose the better suited protocol of the two depending on the specific application scenario. When an address space of $2^{40}$ is sufficient for the in-vehicle traffic of a specific car, then TOUCAN is preferable. For example, this is the case with all control traffic for sensors, which is the main scenario, but may not be the case for firmware upgrades in which the use of ISO-TP standard may be required. The latter, however, is a rare scenario that TOUCAN does not aim to tackle. Other solutions may be used.

\begin{figure*}
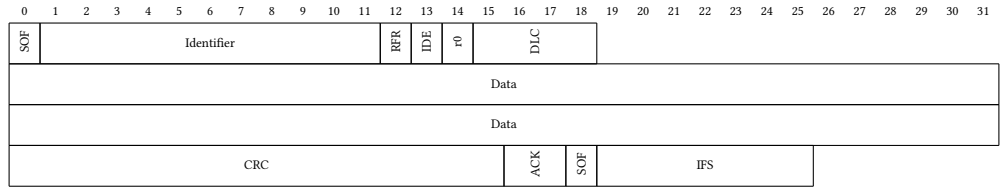

	\begin{minipage}{\textwidth}
		%	%\section*{Appendix}	
		\begin{center}
			%\begin{figure*}[b]
			\begin{bytefield}[bitwidth=1.3em]{32}
				\bitheader{0-31}\\
				\bitbox{1}{\rotatebox[origin=l]{90}{\tiny SOF}}&
				\bitbox{11}{\tiny Identifier} & \bitbox{1}{\rotatebox[origin=l]{90}{\tiny RFR}}&
				\bitbox{1}{\rotatebox[origin=l]{90}{\tiny IDE}} & \bitbox{1}{\rotatebox[origin=l]{90}{\tiny r0}} & \bitbox{4}{\tiny \rotatebox[origin=l]{90}{\tiny DLC}}\\
				\bitbox{32}{\tiny Data}\\
				\bitbox{32}{\tiny Data}\\
				\bitbox{16}{\tiny CRC}&
				\bitbox{2}{\rotatebox[origin=l]{90}{\tiny ACK}} &
				\bitbox{1}{\rotatebox[origin=l]{90}{\tiny SOF}} & \bitbox{7}{\tiny IFS}
			\end{bytefield}
			\captionof{figure}{Standard CAN 2.0A frame format}\label{fig:canframe}
			%\end{figure*} 
		\end{center}
	\end{minipage}
\end{figure*}

\section{Conclusions}\label{sec:concl}
The CAN protocol does not account for the security of the data it handles. By contrast, securing in-vehicle communication is one of the stringent requirements recently introduced by the AUTOSAR standard. This paper presents the design and prototype implementation of TOUCAN, a protocol to secure CAN communication against an active eavesdropper in an AUTOSAR compliant way. It can be deployed by updating the firmware of existing ECUs but demands no hardware upgrade to the network. It is based on fast hashing and symmetric encryption with the aim of ensuring authenticity, integrity and confidentiality. It reduces the payload size to 40 bits but this is largely sufficient for all control traffic. A prototype is implemented on a STM32F407 Discovery board and the runtimes from computing the chosen hashing and encryption functions never exceed a dozen micro seconds.

This paper covered only the operational aspects of secure communication and not its assumptions, the main one being the secure distribution of cryptographic keys that are necessary to bootstrap both the hashing and the encryption primitives. Therefore, future work includes evaluating and relaxing this assumption in practice, also through a reassessment of the runtimes. Future work also sees the simulation of an in-vehicle network by having at least two ECUs communicate securely between each other. Another aspect worth investigating is the precise evaluation of the extent to which more expensive and performing boards than the STM32F407 Discovery used here can reduce the runtimes.

The epoch of \emph{by-design}, secure cars is only just dawning. So far, the literature shows how traditional approaches to network security are being ported to the new setting with the necessary adjustments, mainly due to performance limitations of the ECUs (which can be seen as reactive systems). For example, the gist of all protocols discussed above may be taken to resemble that of IPSec, namely first distribute a symmetric key and then use it to secure the payload. However, on one hand, one may conjecture that performance will not be an issue over time due to the inherent decrease of the cost of computational power. On the other hand, it remains to be seen whether more in-depth assessment followed by more creative thinking and completed with yet more innovative solutions will be necessary. This is the fork that we envisage at the horizon of automotive security.

\bibliographystyle{ACM-Reference-Format}
\bibliography{automotive}

\begin{appendix}

\section{A primer on the CAN bus}\label{sec:can}
CAN frames are standardised as ISO 11898-1:2015~\cite{isocan} to contain various fields, as pictured in Figure~\ref{fig:canframe}, and 
described below.

\begin{itemize}
\item \textbf{Start Of Frame (SOF)} is a dominant bit indicating the beginning of a frame.
\item \textbf{Arbitration field} consists in:
 \emph{Identifier}, 11 bits, to signify the priority of the message, with a lower value indicating higher priority, and \emph{Remote Transmission Request} (RTR), 1 bit, which is low for a Data Frame and high for a Remote Frame (one whose Data Field is empty). 
\item \textbf{Control field}, includes the IDE field, 1 bit, to identify whether the payload is of standard length, then r0, 1 bit, reserved for later use, and the Data Length Code (DLC) field, 4 bits, indicating the length of the Data Field.
\item \textbf{Data} spans over up to 64 bits of data, and carries the payload of the frame.
\item \textbf{CRC}, 15 bits, is for a cyclic redundancy check code and a recessive bit as a delimiter.
\item \textbf{Ack}, 2 bits, with the first one being recessive, hence overwritten with a dominant bit by every node that receives it, and the second bit working as a delimiter.
\item \textbf{EOF}, 7 bits, all recessive, indicates the  end-of-frame.
\item \textbf{IFS}, 7 bits, indicates the time for the controller to move a correct frame into the buffer.\\
\end{itemize}
\vspace*{-4mm}
The mapping between the messages in the payload and the vehicles functionality is up to the car manufacturer and are normally kept confidential. Each mapping enables the ECUs of a specific vehicle to correctly interpret the messages and translate them into signals that carry out the  expected functionality.
\end{appendix}

\end{document}